\documentclass[%
 reprint,
 amsmath,amssymb,
 aps,
]{revtex4-2}

\usepackage{graphicx}
\usepackage{dcolumn}
\usepackage{bm}

\begin{document}

\title{Tachyonic cosmology with cubic contractions of the Riemann tensor}

\author{Mihai Marciu}
 \affiliation{Faculty of Physics, University of Bucharest}
 \email{mihai.marciu@drd.unibuc.ro}

\date{\today}

\begin{abstract}
A new cosmological theory is proposed in the theoretical framework of modified gravity theories which is based on a tachyonic field non--minimally coupled with a specific topological invariant constructed with third order contractions of the Riemann tensor. After proposing the action of the cosmological model, the modified Friedmann relations and the Klein--Gordon equations are obtained, describing the corresponding geometrical corrections to the Einstein--Hilbert action. The physical features of the cosmological system are investigated by adopting the dynamical system analysis in the case of an exponential function for the geometrical coupling term. The investigation revealed that the cosmological system can explain the current accelerated expansion of the Universe and the matter dominated epoch, showing a high compatibility to the recent history of our Universe for various values of the coupling coefficients. 
\end{abstract}

\maketitle

\section{\label{sec:level1}Introduction}

\par 
The general relativity theory represents an important success in describing the behavior of the Universe at the large scale structure dynamics. This theory represents the fundamental structure considered for describing the large scale dynamics of the Universe, later embedded into the $\Lambda$CDM model \cite{Will:2014kxa}. Although the $\Lambda$CDM simple model can explain various phenomena at the large scale \cite{Will:2014kxa}, it suffers from various pathological inconsistencies \cite{Riess:2020fzl, KiDS:2020suj, DiValentino:2021izs, Vagnozzi:2019ezj}. The modified gravity approach \cite{Nojiri:2006ri, Bamba:2012cp, Nojiri:2003ft, Nojiri:2010wj, Nojiri:2017ncd} represents an important direction which aims to further correct the basic theory of general relativity, by taking into account various possible couplings with different specific invariant components which are embedded into the corresponding action \cite{Nojiri:2017ncd}. To this regard, many studies have proposed different modifications of the basic Einstein--Hilbert action, by including or extending the geometrical sector \cite{Nojiri:2017ncd, Frusciante:2019xia}. The main aim of these theories is related to the description of the accelerated expansion at a consistent level, offering some possible solution to various fundamental problems in the modern cosmology \cite{Copeland:2006wr, Bamba:2012cp, Capozziello:2011et}. From an experimental point of view, the accelerated expansion represents a cryptic phenomena at the large scale, affecting the large scale evolution of the Universe \cite{Copeland:2006wr, Planck:2018vyg}. The consistent description and analysis of this phenomena is expected to offer new insights and revolutions in the modern cosmology, having specific effects in the development of various physical theories and technologies. Since the discovery, the accelerated expansion have been probed through various astrophysical studies \cite{Gonzalez-Moran:2021drc, Bermejo-Climent:2021jxf, Cao:2021ldv, Yang:2021flj, Muccino:2020gqt}.
\par 
In the modified gravity landscape the Einsteinian cubic gravity \cite{Bueno:2016xff} represents an interesting and possible viable theory studied in the recent years \cite{Hennigar:2016gkm, Bueno:2016lrh, Hennigar:2018hza, Mehdizadeh:2019qvc, Cisterna:2018tgx, Poshteh:2018wqy, Bueno:2018xqc, Konoplya:2020jgt, Adair:2020vso, Giri:2021amc, Marciu:2021rdl}. This theory is based on a specific invariant, constructed using third order contractions of the Riemann tensor \cite{Bueno:2016xff}. The non--linear extension of the Einsteinian cubic gravity was proposed in Ref.~\cite{Erices:2019mkd}, a theory capable of explaining the late time acceleration of the Universe. The dynamical analysis of the later theory was performed in Refs.~\cite{Marciu:2020ysf, Quiros:2020uhr} for specific parameterizations in the action. Recently, various black hole solutions have been investigated in the Einsteinian cubic gravity theory \cite{Adair:2020vso, KordZangeneh:2020qeg, Frassino:2020zuv, Burger:2019wkq, Cano:2019ozf, Emond:2019crr}. Furthermore, the properties of the inflationary era have been analyzed in some specific theories \cite{Arciniega:2018fxj, Arciniega:2018tnn, Edelstein:2020nhg, Arciniega:2019oxa, Quiros:2020eim, Cano:2020oaa, Edelstein:2020lgv}. The coupling of a scalar field with a cubic term has been investigated by considering a dynamical system analysis \cite{Marciu:2020ski}. Recently, it has been shown that the cubic gravity theory can be associated to the developments of various types of pathological instabilities \cite{Pookkillath:2020iqq} which have to be addressed in order to construct a viable theory  \cite{Jimenez:2020gbw}.
\par 
In the scalar tensor theories, a special class of dark energy models is represented by the tachyonic cosmologies, a novel approach originated from string theory \cite{Mazumdar:2001mm, Sen:2003mv}. The development of tachyonic models has been considered in the past years, leading to various theoretical constructions \cite{Piao:2002vf, Bagla:2002yn, Shchigolev:2012jx, Shchigolev:2011nma, Avelino:2011ey, Gibbons:2002md, Mukohyama:2002cn} which can explain the recent accelerated expansion at the large scale structure. In the scalar tensor theories based on general relativity the study of tachyonic fields have been considered in various cosmological applications \cite{Teixeira:2019tfi, Gibbons:2003gb, Copeland:2004hq, Banijamali:2017lzl, Sen:2002an}. Moreover, the study of cosmological models containing tachyonic fields have been applied in teleparallel gravity \cite{Banijamali:2012nx, Bahamonde:2019gjk, RezaeiAkbarieh:2018ijw, Motavalli:2016gid, Fazlpour:2014qaa, Banijamali:2014nba, Otalora:2013dsa}, a viable alternative theory \cite{Bahamonde:2021gfp}. In these theories, the choice of the potential energy term play a fundamental role, dictating the future and past dynamics of the aforementioned models \cite{Abramo:2003cp, Copeland:2004hq}. The study of tachyonic cosmological scenarios for various classes of potential energies have been considered \cite{Quiros:2009mz, Fang:2010zze, Bahamonde:2019gjk, Abramo:2003cp}, leading to viable models which can explain various cosmological features.

\par 
In the present paper we shall further extend the tachyonic cosmology by considering a non--minimal coupling with a novel topological invariant, based on cubic contractions of the Riemann tensor. After we deduce the corresponding field equation, we shall study the physical implications by adopting the dynamical system analysis. The study takes into consideration two specific cases associated to the behavior of the potential energy term. In the first case we consider an exponential representation, while in the second one an inverse hyperbolic sine function is studied. For all of the previous mentioned cases we have analyzed the structure and properties of the phase space, discussing possible physical effects. 

\par 
The present manuscript is organized as follows: in Sec.~\ref{sec:level2} we discuss the action and the corresponding field equations for the dark energy model. Then, in Sec.~\ref{sec:level3} we analyze the physical features for an exponential coupling and potential by considering the dynamical system analysis. In Sec.~\ref{sec:level4aa} we discuss the phase space structure where the potential energy term is beyond the usual exponential case, considering an inverse hyperbolic function. Finally, in Sec.~\ref{sec:level5} we have a short summary of our analysis, discussing the main conclusions which are applicable to the present study.

\section{\label{sec:level2}The description of the field equations}
\par 
In the present study we shall consider a tachyonic cosmological model non--minimally coupled with a topological invariant constructed from the cubic contractions of the Riemann tensor. The action corresponding to the present study is the following:

\begin{equation}
\label{eq:actiune}
    S=S_m+\int d^4 x \sqrt{-g} \Big[ \frac{R}{2} - V(\phi) \sqrt{1+\epsilon \frac{g^{\mu \nu}\partial_{\mu} \phi \partial_{\nu} \phi}{V(\phi)}} + f(\phi) P \Big], 
\end{equation}
where the topological invariant is based on specific contractions of the Riemann tensor in the third order \cite{Erices:2019mkd}, 

\begin{multline}
P=\beta_1 R_{\mu\quad\nu}^{\quad\rho\quad\sigma}R_{ \rho\quad\sigma}^{\quad \gamma\quad\delta}R_{\gamma\quad\delta}^{\quad\mu\quad\nu}+\beta_2 R_{\mu\nu}^{\rho\sigma}R_{\rho\sigma}^{\gamma\delta}R_{\gamma\delta}^{\mu\nu}
\\+\beta_3 R^{\sigma\gamma}R_{\mu\nu\rho\sigma}R_{\quad\quad\gamma}^{\mu\nu\rho}+\beta_4 R R_{\mu\nu\rho\sigma}R^{\mu\nu\rho\sigma}+\beta_5 R_{\mu\nu\rho\sigma}R^{\mu\rho}R^{\nu\sigma}
\\+\beta_6 R_{\mu}^{\nu}R_{\nu}^{\rho}R_{\rho}^{\mu}+\beta_7 R_{\mu\nu}R^{\mu\nu}R+\beta_8 R^3,
\end{multline} 
with $\beta_{j}, (j=\{1,...,8\})$ constant parameters.
\par 
In this case the potential energy is $V(\phi)$, and $\epsilon$ is a constant parameter which describes the canonical representation of the tachyonic field. For a canonical tachyonic field $\epsilon=+1$, while for the non--canonical case we have $\epsilon=-1$. The matter part in the action is denoted by $S_m$, describing a barotropic fluid which characterizes the dark matter sector, having the corresponding density $\rho_m$ and pressure $p_m$, satisfying the equation of state $p_m=\rho_m w_m$, with $w_m$ a constant coefficient describing a non--relativistic behavior. In this case, the dark matter fluid satisfies the standard continuity equation. 
\par 
Next, the large scale structure dynamics in the Universe is described by the following Robertson--Walker metric, 
\begin{equation}
\label{metrica}
ds^2=-dt^2+a^2(t) \delta_{u v}dx^u dx^v,
\end{equation}
with $a(t)$ the scale factor, $H(t)$ the Hubble parameter, and $t$ the cosmic time. If we adopt the following relations between the constant parameters $\beta_{j}, (j=\{1,...,8\})$ \cite{Erices:2019mkd}, 
\begin{equation}
\beta_7=\frac{1}{12}\big[3\beta_1-24\beta_2-16\beta_3-48\beta_4-5\beta_5-9\beta_6\big],
\end{equation}
\begin{equation}
\beta_8=\frac{1}{72}\big[-6\beta_1+36\beta_2+22\beta_3+64\beta_4+3\beta_5+9\beta_6\big],
\end{equation}
\begin{equation}
\beta_6=4\beta_2+2\beta_3+8\beta_4+\beta_5,
\end{equation}
\begin{equation}
\bar{\beta}=(-\beta_1+4\beta_2+2\beta_3+8\beta_4),
\end{equation}
then the third order tensor polynomial constructed from the cubic contractions of the Riemann tensor is equal to \cite{Erices:2019mkd}:
\begin{equation}
\label{PP}
P=6\bar{\beta}H^4 (2H^2+3\dot{H}).
\end{equation}
\par
Here, the dot represents the differentiation with respect to the cosmic time $t$, while the prime denotes the differentiation with respect to the argument of the specific function. The variation of the action described in the Eq.~\ref{eq:actiune} with respect to the tachyonic field $\phi(t)$ gives the corresponding Klein--Gordon equation, 
\begin{multline}
\label{klein}
\Big[1-\epsilon \frac{\dot{\phi}^2}{V(\phi)} \Big]^{-\frac{3}{2}} \Big[ 3 \epsilon \frac{V'(\phi)\dot{\phi}^2}{V(\phi)}-2 (V'(\phi)+\epsilon \ddot{\phi}) \Big]\\ + 
12 \beta H^4 f'(\phi) (2 H^2+3 \dot{H})-\frac{6 \epsilon H \dot{\phi}}{\sqrt{1-\epsilon \frac{\dot{\phi}^2}{V(\phi)}}}=0.
\end{multline}
\par 
Furthermore, the variation of the action \eqref{eq:actiune} with respect to the inverse metric leads to the modified Friedmann relations which have the following form \cite{Erices:2019mkd}:

\begin{equation}
\label{friedmannconstr}
3H^2=\rho_m+\rho_{\phi},
\end{equation}
\begin{equation}
\label{friedmannaccelerare}
3H^2+2\dot{H}=-p_m-p_{\phi},
\end{equation}
where the energy density of the tachyonic field is  
\begin{equation}
\label{densitatede}
\rho_{\phi}=\frac{V(\phi)}{\sqrt{1-\epsilon \frac{\dot{\phi}^2}{V(\phi)}}}+6 \beta f(\phi) H^6-18 \beta H^5 \frac{df(\phi)}{d\phi}\dot{\phi},
\end{equation}
with the pressure
\begin{multline}
\label{presiunede}
p_{\phi}=-V(\phi)\sqrt{1-\epsilon \frac{\dot{\phi}^2}{V(\phi)}}-6 \beta f(\phi) H^6-12 \beta f(\phi) H^4 \dot{H}
\\+12 \beta H^5 \frac{df(\phi)}{d\phi}\dot{\phi}+24 \beta H^3 \frac{df(\phi)}{d\phi}\dot{H}\dot{\phi}
\\
+6 \beta H^4 \dot{\phi}^2\frac{d^2f(\phi)}{d\phi^2}+6 \beta H^4 \frac{df(\phi)}{d\phi}\ddot{\phi}.
\end{multline}
\par 
Finally, we can define the barotropic parameter associated to the dark energy field,
\begin{equation}
w_{\bf{\phi}}=\frac{p_{\phi}}{\rho_{\phi}},
\end{equation}
and the effective (total) equation of state for our cosmological model, 

\begin{equation}
w_{\bf{eff}}=\frac{p_m+p_{\phi}}{\rho_{m}+\rho_{\phi}}=-1-\frac{2}{3}\frac{\dot{H}}{H^2}.
\end{equation}
\par 
If we introduce the matter density parameter,
\begin{equation}
\Omega_m=\frac{\rho_{m}}{3H^2},
\end{equation}
and the density parameter corresponding to the tachyonic field,
\begin{equation}
\Omega_{\phi}=\frac{\rho_{\phi}}{3H^2},
\end{equation}
we have the following constraint,
\begin{equation}
\Omega_m+\Omega_{\phi}=1.
\end{equation}

\section{\label{sec:level3}Dynamical effects in the case of an exponential potential}

\par 
In this section we shall discuss the main physical features of the present cosmological model by applying the dynamical system analysis, an important tool in the study of various modified gravity theories. Analyzing the Friedmann constraint equation \eqref{friedmannconstr} we introduce the following dimension--less variables:

\begin{equation}
  x=\frac{\dot{\phi}}{\sqrt{V(\phi)}},  
\end{equation}

\begin{equation}
  y=\frac{\sqrt{V(\phi)}}{H\sqrt{3}},  
\end{equation}

\begin{equation}
    z=\beta f(\phi) H^4,
\end{equation}

in the case of an exponential coupling, 

\begin{equation}
    f(\phi)=f_0 e^{\alpha \phi},
\end{equation}

and potential energy,  

\begin{equation}
    V(\phi)=e^{-\lambda \phi}.
\end{equation}
\par 
Hence, in this case we have
\begin{equation}
    V'(\phi)=-\lambda V(\phi),
\end{equation}
where $\lambda$ is a positive constant coefficient which characterizes the steepness of the potential energy. \par 
The Friedmann constraint equation \eqref{friedmannconstr} can be written as:
\begin{equation}
    1=\frac{y^2}{\sqrt{1-\epsilon x^2}}+2 z -6 \alpha z x y \sqrt{3}+\Omega_m.
\end{equation}

Next, if we introduce another variable $N=log(a)$ and change the dependence of the dimension--less components to $N$ (the e-fold variable), we obtain the following autonomous dynamical system, approximating the evolution of the cosmological system: 

\begin{equation}
    x'=\frac{\ddot{\phi}}{H^2} \frac{1}{y \sqrt{3}}+\frac{\lambda}{2} \sqrt{3} x^2 y,
\end{equation}

\begin{equation}
    y'=-\lambda \frac{\sqrt{3}}{2} x y^2-y \frac{\dot{H}}{H^2},
\end{equation}

\begin{equation}
    z'=\alpha z x y \sqrt{3} + 4 z \frac{\dot{H}}{H^2},
\end{equation}
where the prime $'$ describe the derivative with respect to $N$.

\par 
In this case the Klein--Gordon equation \eqref{klein} can be written in the following way:
\onecolumngrid
\begin{equation}
    -\frac{6 \sqrt{3} H^2 x y \epsilon }{\sqrt{1-x^2 \epsilon }}+\frac{-27 H^4 \lambda  x^2 y^4 \epsilon -6 H^2 y^2 \left(\epsilon  \ddot{\phi}-3 H^2 \lambda  y^2\right)}{3 H^2 y^2 \left(1-x^2 \epsilon \right)^{3/2}}+12 \alpha  z \left(2 H^2+3 \dot{H}\right)=0,
\end{equation}
while the acceleration equation \eqref{friedmannaccelerare} is equal to:
\begin{equation}
    -3 H^2-2 \dot{H}=3 H^2 w_m \Omega_m+18 \alpha ^2 H^2 x^2 y^2 z-3 H^2 y^2 \sqrt{1-x^2 \epsilon }+12 \sqrt{3} \alpha  H^2 x y z-6 H^2 z+24 \sqrt{3} \alpha  x y z \dot{H}-12 z \dot{H}+6 \alpha  z \ddot{\phi}.
\end{equation}

\par 
In this way we obtain the final expression of the autonomous dynamical system:
\begin{multline}
    x'=\frac{1}{y \left(54 \alpha ^2 z^2 \left(1-x^2 \epsilon \right)^{3/2}+6 z \epsilon  \left(2 \sqrt{3} \alpha  x y-1\right)+\epsilon \right)} \cdot 
    \\ \Big(-9 \sqrt{3} \alpha  x^2 y^2 z \epsilon  w_m-162 \alpha ^2 x y z^2 w_m \sqrt{1-x^2 \epsilon }-18 \sqrt{3} \alpha  x^2 z^2 \epsilon  w_m \sqrt{1-x^2 \epsilon }+18 \sqrt{3} \alpha  z^2 w_m \sqrt{1-x^2 \epsilon }+9 \sqrt{3} \alpha  x^2 z \epsilon  w_m \sqrt{1-x^2 \epsilon }
    \\-9 \sqrt{3} \alpha  z w_m \sqrt{1-x^2 \epsilon }+162 \alpha ^2 x^3 y z^2 \epsilon  w_m \sqrt{1-x^2 \epsilon }+9 \sqrt{3} \alpha  y^2 z w_m+45 \sqrt{3} \alpha  x^4 y^2 z \epsilon ^2-36 \alpha  \lambda  x^3 y^3 z \epsilon -18 x^3 y z \epsilon ^2+3 x^3 y \epsilon ^2
    \\-54 \sqrt{3} \alpha ^3 x^2 y^2 z^2 \sqrt{1-x^2 \epsilon }+27 \sqrt{3} \alpha ^2 \lambda  x^2 y^2 z^2 \sqrt{1-x^2 \epsilon }
    -54 \sqrt{3} \alpha  x^2 y^2 z \epsilon +6 \sqrt{3} \lambda  x^2 y^2 z \epsilon -\sqrt{3} \lambda  x^2 y^2 \epsilon +36 \alpha ^2 x y z^2 \sqrt{1-x^2 \epsilon }
    \\+6 \sqrt{3} \alpha  x^2 z^2 \epsilon  \sqrt{1-x^2 \epsilon }-6 \sqrt{3} \alpha  z^2 \sqrt{1-x^2 \epsilon }
    +5 \sqrt{3} \alpha  x^2 z \epsilon  \sqrt{1-x^2 \epsilon }
    -5 \sqrt{3} \alpha  z \sqrt{1-x^2 \epsilon }+54 \sqrt{3} \alpha ^3 x^4 y^2 z^2 \epsilon  \sqrt{1-x^2 \epsilon }
    \\-27 \sqrt{3} \alpha ^2 \lambda  x^4 y^2 z^2 \epsilon  \sqrt{1-x^2 \epsilon }-36 \alpha ^2 x^3 y z^2 \epsilon  \sqrt{1-x^2 \epsilon }+36 \alpha  \lambda  x y^3 z+18 x y z \epsilon -3 x y \epsilon 
    +\sqrt{3} \lambda  y^2+9 \sqrt{3} \alpha  y^2 z-6 \sqrt{3} \lambda  y^2 z \Big),
\end{multline}

\begin{multline}
y'=\frac{1}{2 \sqrt{1-x^2 \epsilon } \left(54 \alpha ^2 z^2 \left(1-x^2 \epsilon \right)^{3/2}+6 z \epsilon  \left(2 \sqrt{3} \alpha  x y-1\right)+\epsilon \right)} \cdot
\\ \Big( 18 \sqrt{3} \alpha  x y^2 z \epsilon  w_m \sqrt{1-x^2 \epsilon }-6 y z \epsilon  w_m \sqrt{1-x^2 \epsilon }+3 y \epsilon  w_m \sqrt{1-x^2 \epsilon }-3 y^3 \epsilon  w_m-54 \sqrt{3} \alpha ^2 \lambda  x^5 y^2 z^2 \epsilon ^2+72 \alpha ^2 x^4 y z^2 \epsilon ^2
\\+108 \sqrt{3} \alpha ^2 \lambda  x^3 y^2 z^2 \epsilon +18 \alpha ^2 x^2 y^3 z \epsilon  \sqrt{1-x^2 \epsilon }-63 \alpha  \lambda  x^2 y^3 z \epsilon  \sqrt{1-x^2 \epsilon }+18 \alpha  \lambda  y^3 z \sqrt{1-x^2 \epsilon }+3 x^2 y^3 \epsilon ^2-6 \sqrt{3} \alpha  x y^2 z \epsilon  \sqrt{1-x^2 \epsilon }
\\+6 \sqrt{3} \lambda  x y^2 z \epsilon  \sqrt{1-x^2 \epsilon }-\sqrt{3} \lambda  x y^2 \epsilon  \sqrt{1-x^2 \epsilon }-144 \alpha ^2 x^2 y z^2 \epsilon -6 y z \epsilon  \sqrt{1-x^2 \epsilon }+3 y \epsilon  \sqrt{1-x^2 \epsilon }+18 \sqrt{3} \alpha  x^3 y^2 z \epsilon ^2 \sqrt{1-x^2 \epsilon }
\\-54 \sqrt{3} \alpha ^2 \lambda  x y^2 z^2-3 y^3 \epsilon +72 \alpha ^2 y z^2 \Big), 
\end{multline}

\begin{multline}
    z'=\frac{1}{\sqrt{1-x^2 \epsilon } \left(54 \alpha ^2 z^2 \left(1-x^2 \epsilon \right)^{3/2}+6 z \epsilon  \left(2 \sqrt{3} \alpha  x y-1\right)+\epsilon \right)} \cdot \\ \Big( -36 \sqrt{3} \alpha  x y z^2 \epsilon  w_m \sqrt{1-x^2 \epsilon }+12 z^2 \epsilon  w_m \sqrt{1-x^2 \epsilon }-6 z \epsilon  w_m \sqrt{1-x^2 \epsilon }+6 y^2 z \epsilon  w_m+54 \sqrt{3} \alpha ^3 x^5 y z^3 \epsilon ^2-144 \alpha ^2 x^4 z^3 \epsilon ^2
    \\-108 \sqrt{3} \alpha ^3 x^3 y z^3 \epsilon +54 \alpha  \lambda  x^2 y^2 z^2 \epsilon  \sqrt{1-x^2 \epsilon }-36 \alpha  \lambda  y^2 z^2 \sqrt{1-x^2 \epsilon }-6 x^2 y^2 z \epsilon ^2+6 \sqrt{3} \alpha  x y z^2 \epsilon  \sqrt{1-x^2 \epsilon }+\sqrt{3} \alpha  x y z \epsilon  \sqrt{1-x^2 \epsilon }
    \\+288 \alpha ^2 x^2 z^3 \epsilon +12 z^2 \epsilon  \sqrt{1-x^2 \epsilon }-6 z \epsilon  \sqrt{1-x^2 \epsilon }-36 \sqrt{3} \alpha  x^3 y z^2 \epsilon ^2 \sqrt{1-x^2 \epsilon }+54 \sqrt{3} \alpha ^3 x y z^3+6 y^2 z \epsilon -144 \alpha ^2 z^3 \Big).
\end{multline}

\begin{figure}[htb]
    \centering
    \includegraphics[width=0.4\textwidth]{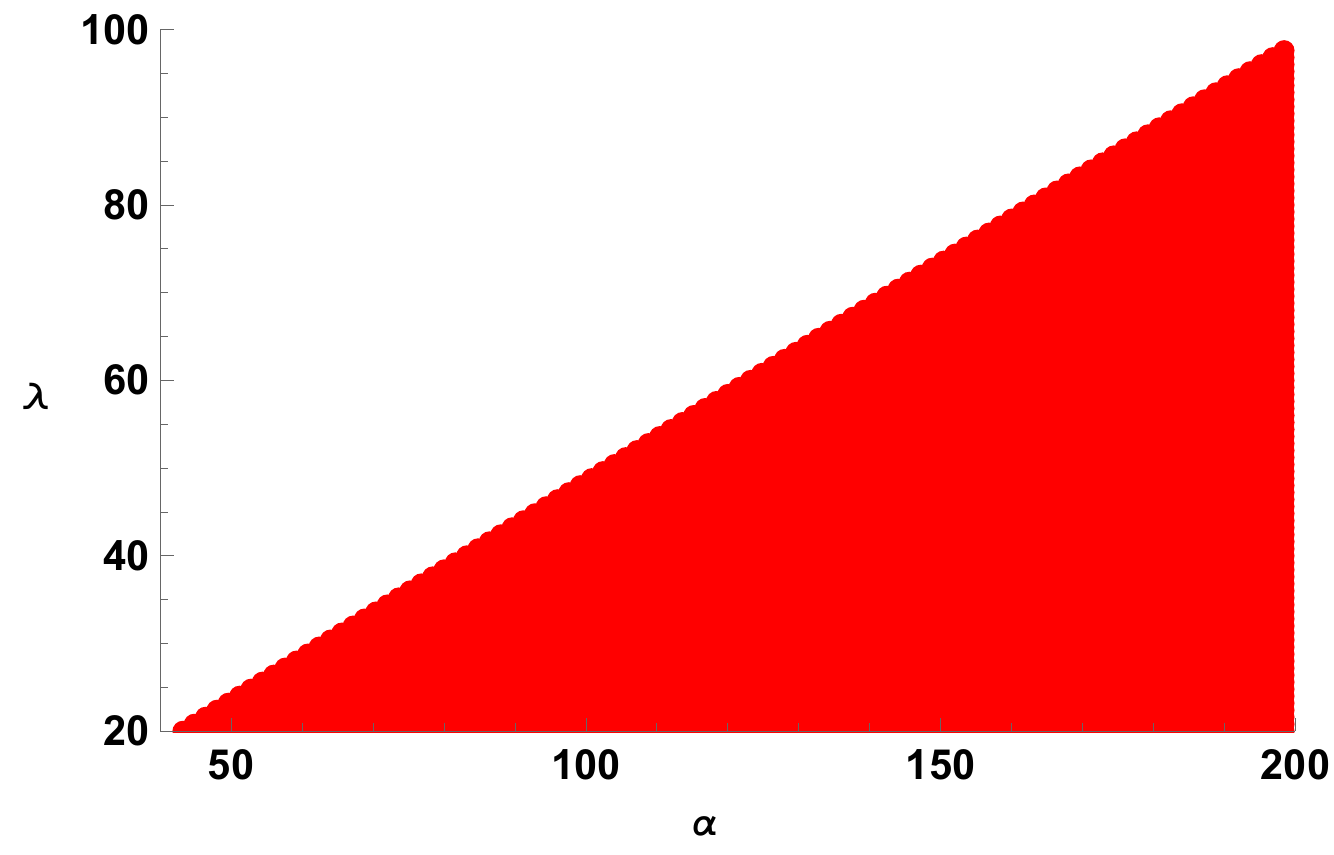}
    \caption{The figure describes a region in the phase space structure where the A cosmological solution is saddle, in the case where $w_m=-0.0001$.}
    \label{fig:asaddle2}
\end{figure}

\begin{figure}[htb]
    \centering
    \includegraphics[width=0.4\textwidth]{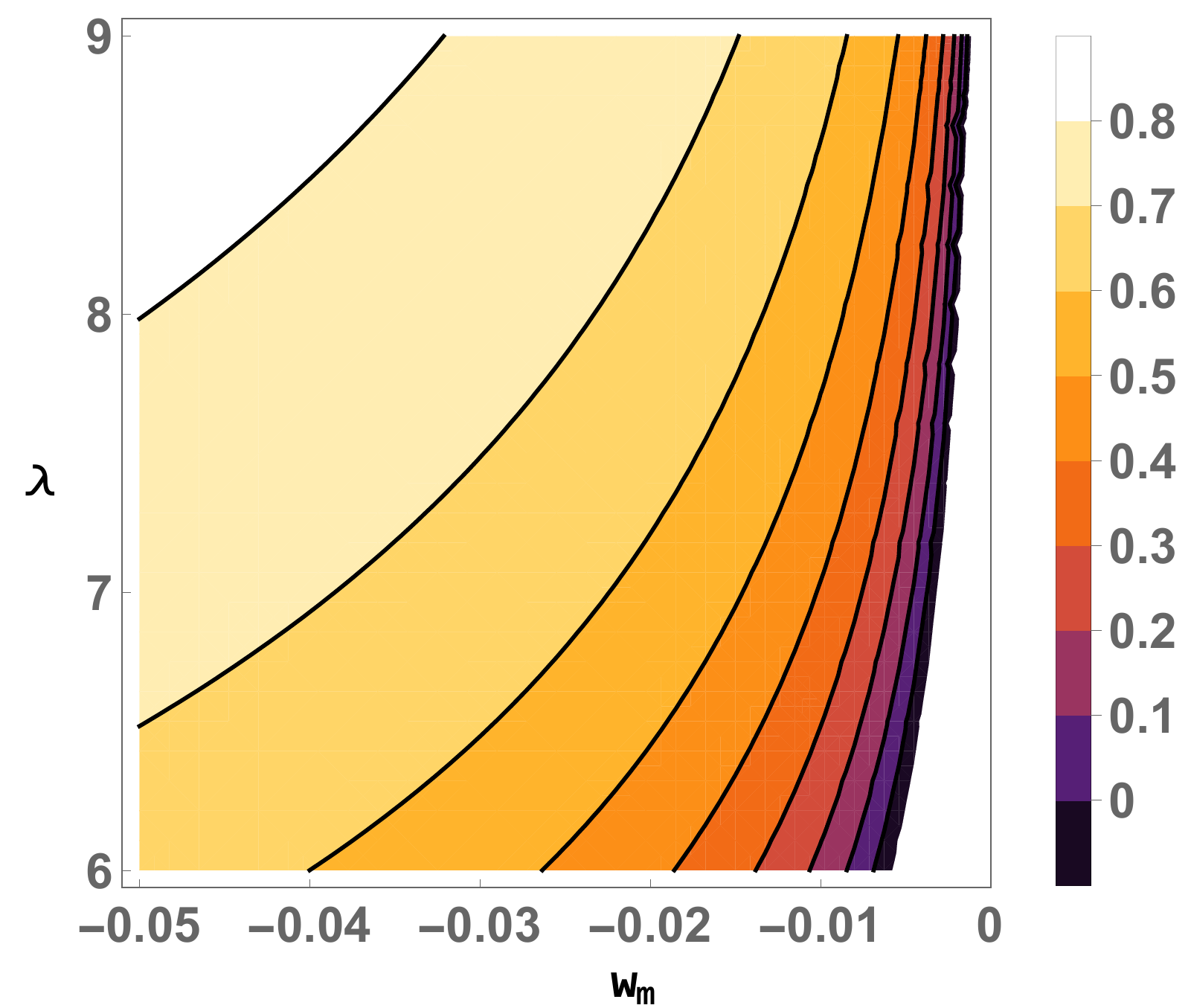}
    \caption{The figure describes the value of the matter density parameter $\Omega_m$ as a function of dark matter equation of state parameter $w_m$ and the strength of the potential energy, encoded into the value of the $\lambda$ coefficient.}
    \label{fig:aomegamatter1}
\end{figure}

\begin{figure}[htb]
    \centering
    \includegraphics[width=0.4\textwidth]{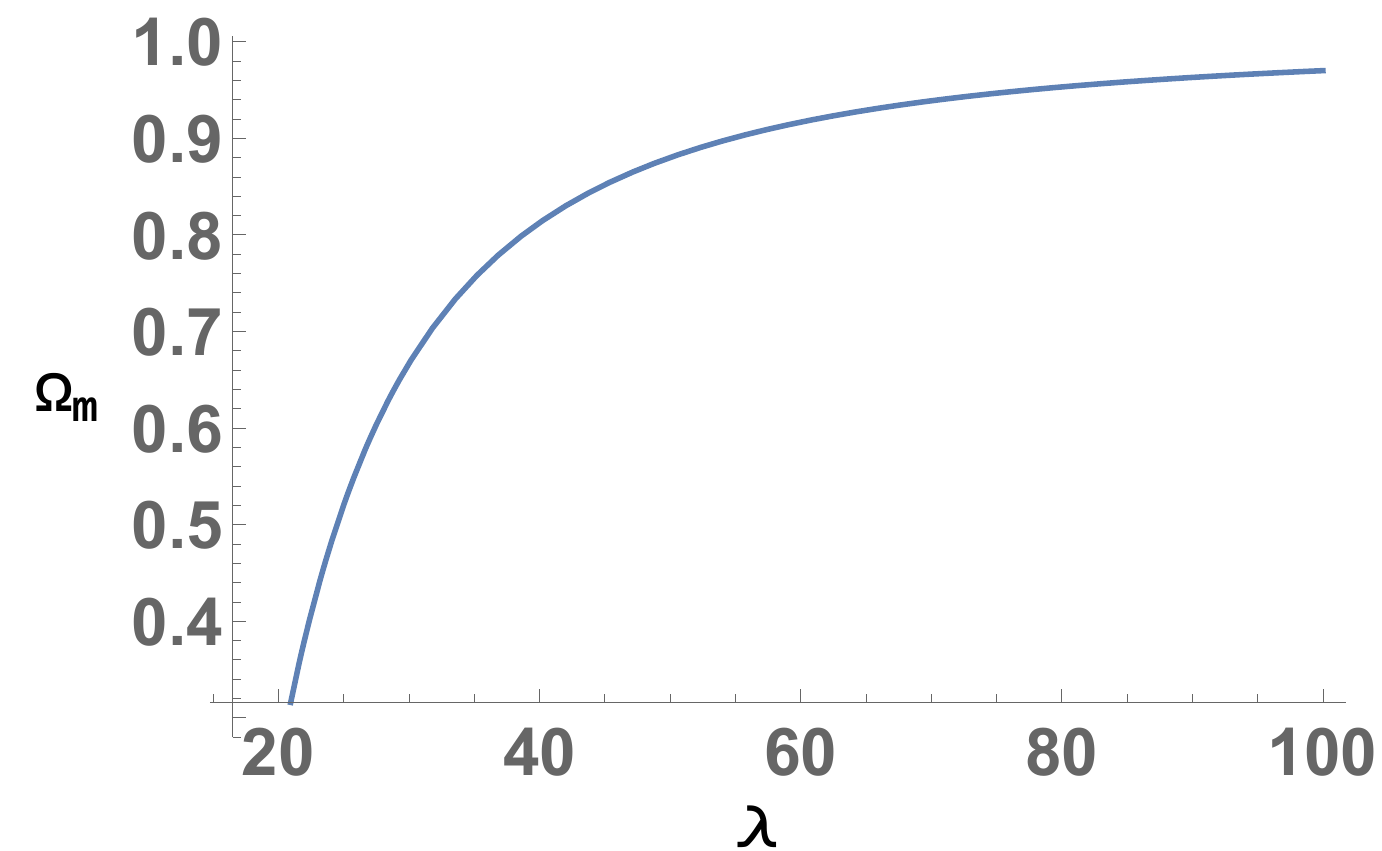}
    \caption{The figure describes the value of the matter density parameter $\Omega_m$ as a function of the strength of the potential energy, encoded into the value of the $\lambda$ coefficient. In this figure we have considered $w_m=-0.0001$.}
    \label{fig:aomegamatter2}
\end{figure}

\begin{figure}[htb]
    \centering
    \includegraphics[width=0.4\textwidth]{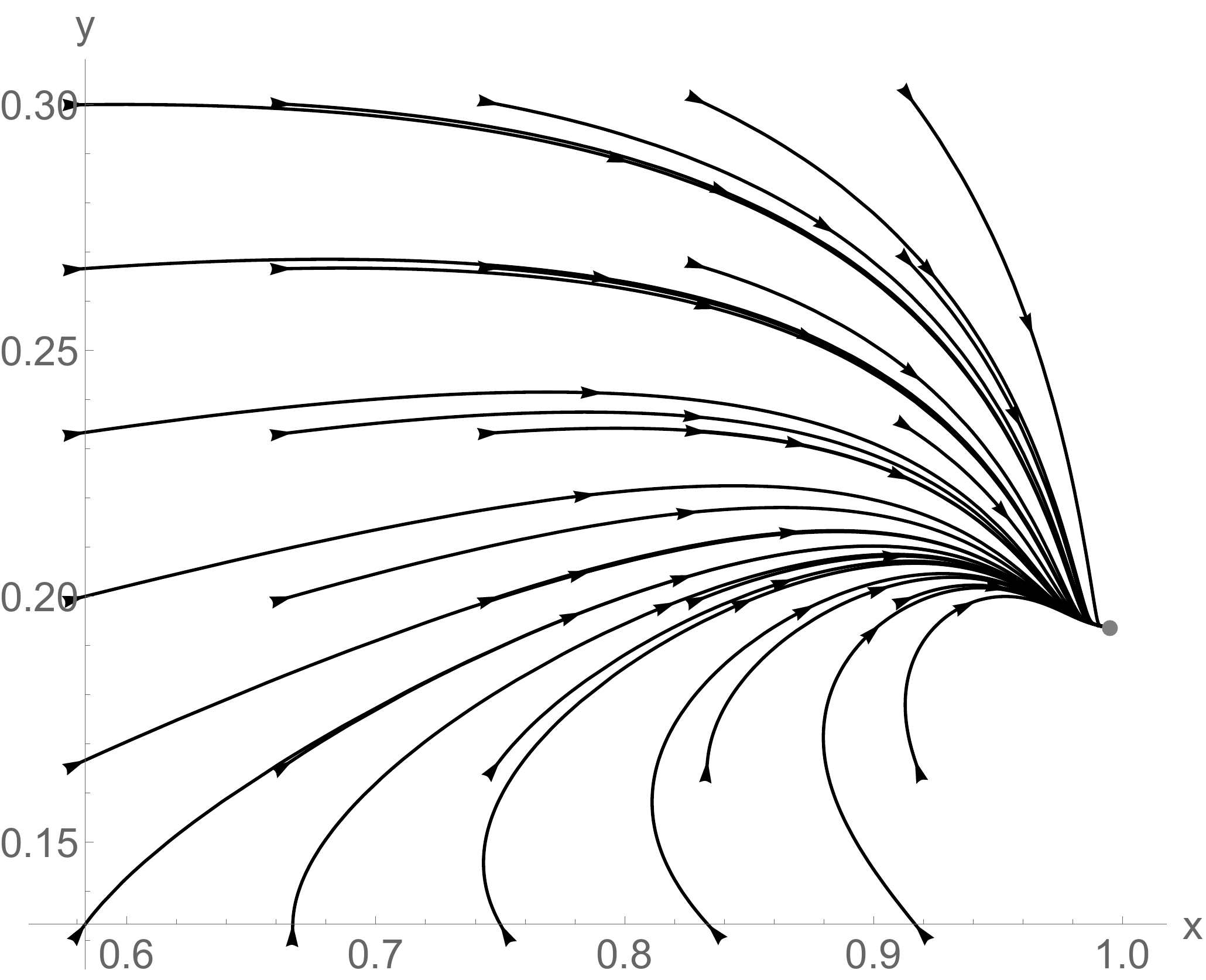}
    \caption{The figure describes the evolution towards the A critical point in the $\{x, y\}$ plane ($\alpha=1, w_m=-0.01, \lambda=8.9, \epsilon=1$).}
    \label{fig:evolutie1}
\end{figure}

\begin{figure}[htb]
    \centering
    \includegraphics[width=0.4\textwidth]{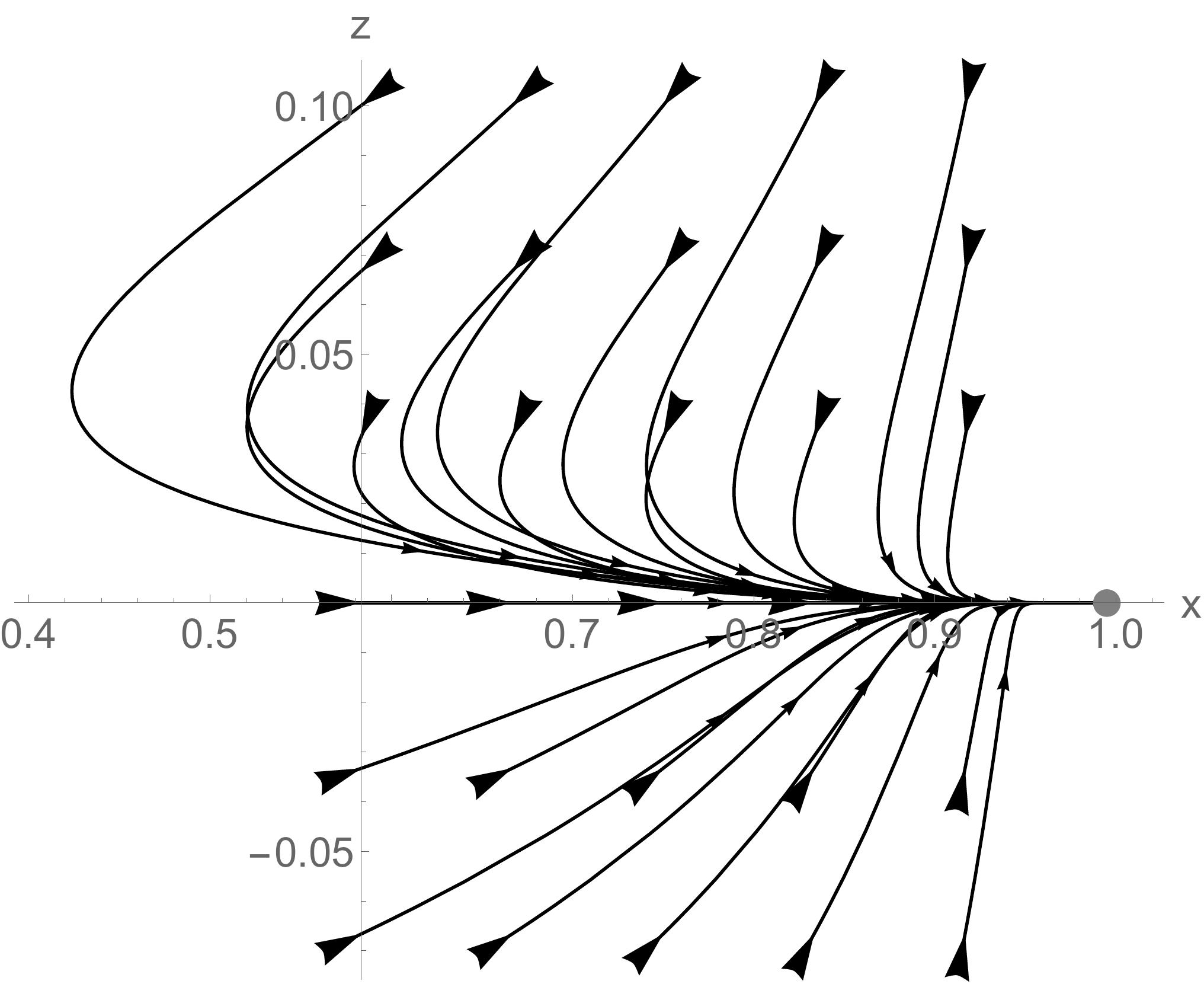}
    \caption{The evolution towards the A cosmological solution in the $\{x, z\}$ plane.}
    \label{fig:evoluti2}
\end{figure}

\begin{figure}[htb]
    \centering
    \includegraphics[width=0.4\textwidth]{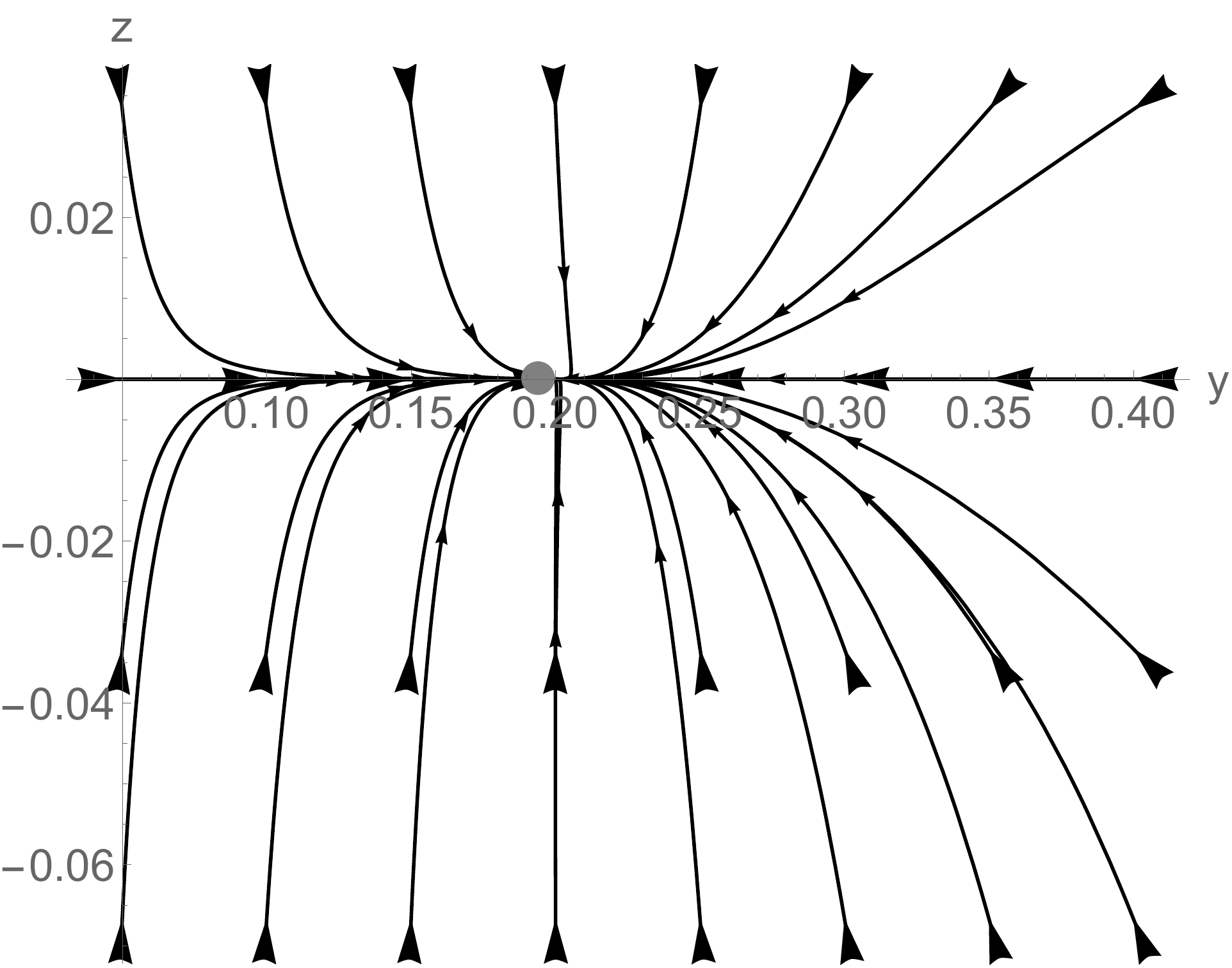}
    \caption{The evolution towards A in the $\{y, z\}$ plane}
    \label{fig:evolutie3}
\end{figure}

\begin{figure}[h!]
    \centering
    \includegraphics[trim={8cm 0cm 0cm 0},clip, width=1.2\textwidth]{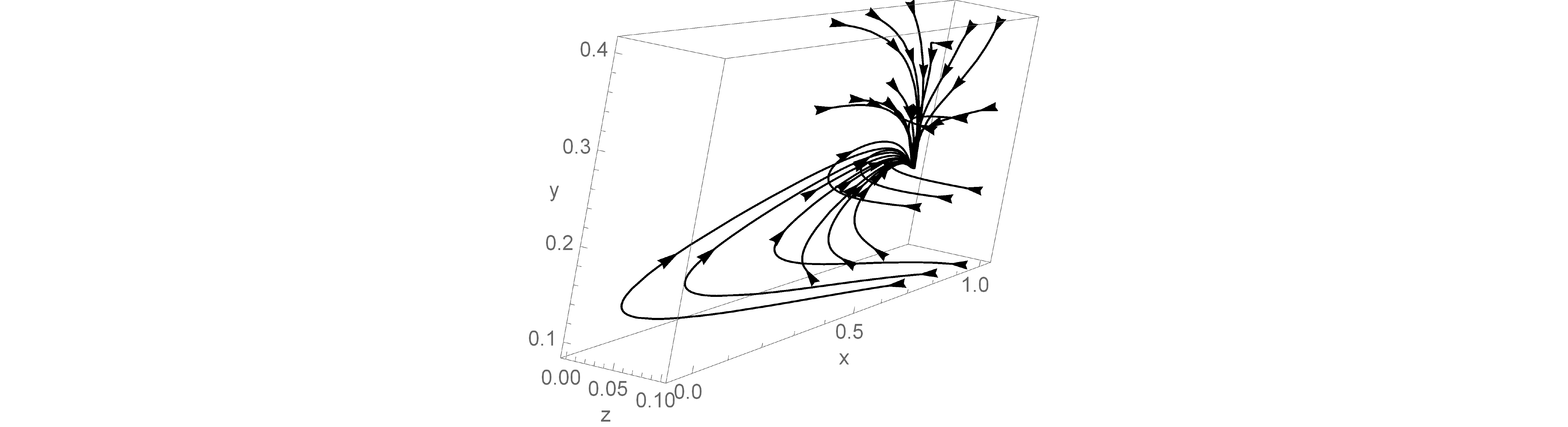}
    \caption{The figure describes the evolution in the 3D space for various initial conditions, towards the A cosmological solution. }
    \label{fig:evolutie4}
\end{figure}

\begin{figure}[h!]
    \centering
    \includegraphics[width=0.4\textwidth]{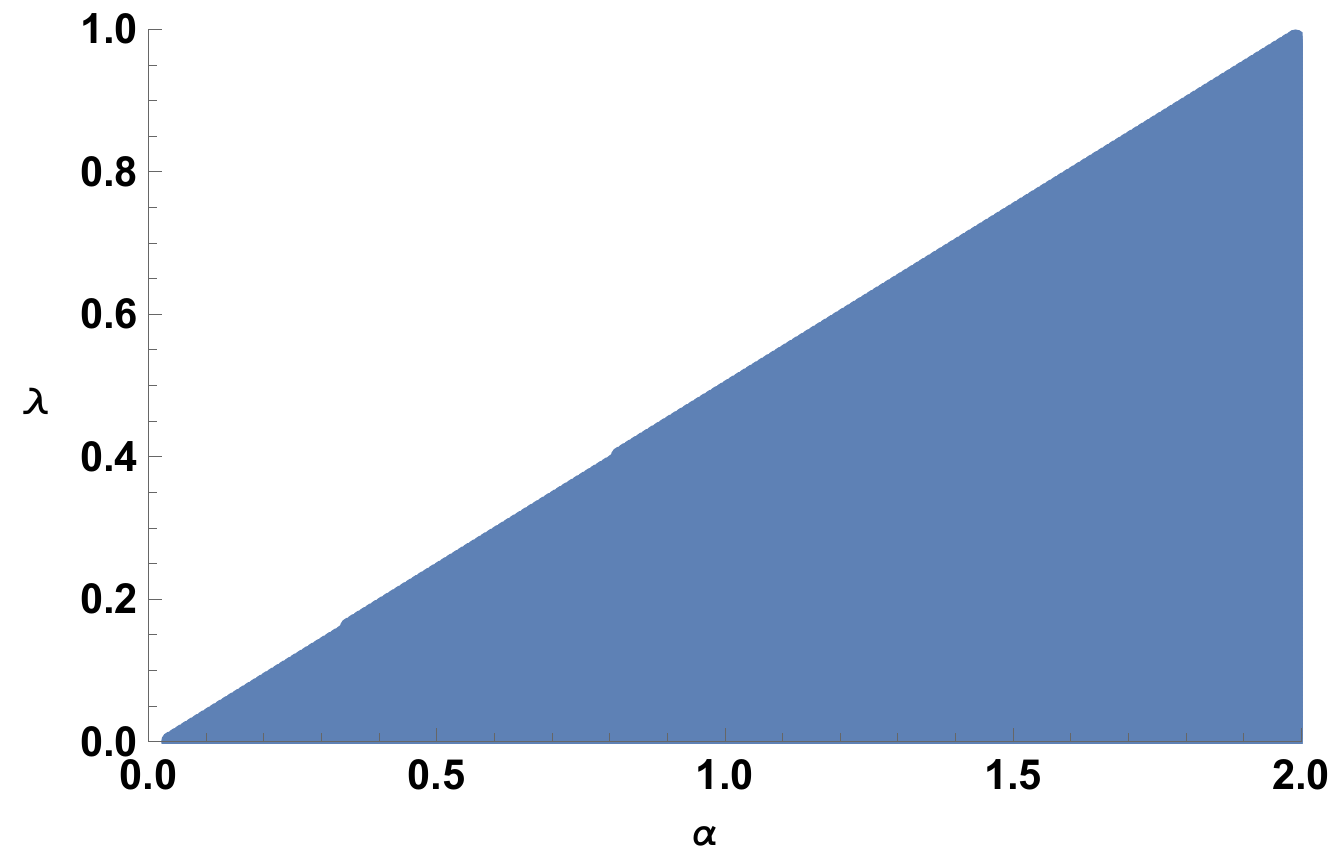}
    \caption{The figure describes a specific region where the de--Sitter critical point B is stable. The figure takes into account also the existence conditions, implying that the $y$ component is real and positive.}
    \label{fig:punctbeiv}
\end{figure}
\twocolumngrid

\par 
For our cosmological system we have obtained two classes of critical points. The first critical point is located at the following coordinates:
\begin{equation}
    A=\Big[x=\sqrt{1+w_m}, y=\frac{\sqrt{3}\sqrt{1+w_m}}{\lambda}, z=0\Big].
\end{equation}
At this solution we note that the dark matter equation of state dictates the value of the kinetic term, while for the potential component we have an influence due to the steepness of the potential, embedded into the value of the $\lambda$ coefficient. The effective equation of state is equal to 
\begin{equation}
    w_{\bf{eff}}=w_m,
\end{equation} with the matter density parameter
\begin{equation}
    \Omega_m=1-\frac{3(1+w_m)}{\sqrt{-w_m}\lambda^2}.
\end{equation} Due to the existence conditions the matter density parameter should be real in the $[0,1]$ interval, implying that the $w_m$ parameter is slightly negative and close to zero, a value not ruled out by astrophysical observations. This cosmological solution represents a matter dominated era, a critical point which appeared also in various dynamical studies \cite{Bahamonde:2019gjk}. This cosmological solution is similar to the one found in the minimal coupling case \cite{Quiros:2009mz}. For this cosmological solution we have obtained the following eigenvalues:
\onecolumngrid
\begin{equation}
    \Big[\frac{3 (\alpha -2 \lambda ) \left(w_m+1\right)}{\lambda }, \frac{3}{4}\left(w_m-1 \pm \frac{\Pi}{\Psi} \right) \Big],
\end{equation}
where 

\begin{equation}
    \Pi=\sqrt{-\lambda ^4 w_m^3 \left(w_m+1\right) \left(\lambda ^2+17 \lambda ^2 w_m^2+14 \lambda ^2 w_m+48 \left(-w_m\right){}^{5/2}+48 \sqrt{-w_m}+96 \sqrt{-w_m} w_m\right)},
\end{equation}
\begin{equation}
    \Psi=\lambda ^3 w_m \sqrt{-w_m \left(w_m+1\right)}.
\end{equation}
\twocolumngrid

From the above expressions we can note that this solution has a high sensitivity to the values of various coefficients, $w_m, \lambda, \alpha$. For this solution, we have displayed in Fig.~\ref{fig:asaddle2} some possible intervals where the corresponding behavior is associated to a saddle dynamics. In the implementation of this figure we have used an interpolation method based on a neural network approach, with an initial grid considered as the training set. The architecture of the neural network is based on four internal linear layers, separated through a specific activation function. After training, the neural network can be considered for the interpolation, obtaining possible regions in the phase space structure where the $A$ cosmological solution has a saddle behavior. For example, in the case where $\Omega_m \approx 0.70$, if we set $\lambda=32$, $w_m=-0.0001$, and $\alpha=100$, we obtain a saddle cosmological solution which can mimic the matter dominated era.
\par 
In Fig.~\ref{fig:aomegamatter1} we have the dependence of the matter density parameter $\Omega_m$ in the $\{ w_m, \lambda \}$ plane. It can be seen that the matter density parameter can span easily by fine--tuning the $[0,1]$ interval, being compatible to various astrophysical observations. Furthermore, in Fig.~\ref{fig:aomegamatter2} we have displayed the value of the matter density parameter $\Omega_m$ as a function of $\lambda$, the strength of the potential energy. In this case we have fixed $w_m=-0.0001$. Lastly, for this solution we have analyzed the phase space structure, displaying various aspects of the numerical evolution in Figs.~\ref{fig:evolutie1}--\ref{fig:evolutie4}.
\par 
The second cosmological solution is located at the coordinates:
\begin{equation}
    B=\Big[x=0, y=\frac{\sqrt{2 \alpha}}{\sqrt{2 \alpha - \lambda}}, z=\frac{\lambda}{2(\lambda-2 \alpha)} \Big], 
\end{equation}
a de--Sitter epoch where the effective equation of state corresponds to a cosmological constant, 
\begin{equation}
    w_{\bf{eff}}=-1.
\end{equation}
\par
The matter density parameter is $\Omega_m=0$, a solution where the scalar field completely dominates in terms of density parameters. For this solution, we can note that the field is at rest, without any kinetic energy. The potential energy variable $y$ and the coupling component which is encoded into the $z$ variable are both influenced by the strength of the coupling function, displayed by the $\alpha$ coefficient, and the steepness of the exponential potential. From a dynamical point of view we have obtained the following eigenvalues, 

\begin{equation}
    \Bigg[-3 \left(w_m+1\right), \frac{1}{\Theta} (\Delta \pm \sqrt{3} \sqrt{\Xi})   \Bigg],
\end{equation}
 where
 \onecolumngrid
 
 \begin{equation}
     \Theta=2 (2 \alpha -\lambda )^3 \left(\alpha ^2 \left(27 \lambda ^2+8\right)+4 \alpha  \lambda -4 \lambda ^2\right)^2,
 \end{equation}
 
 \begin{multline}
     \Delta=24 \alpha ^7 \left(27 \lambda ^2+8\right)^2-12 \alpha ^6 \lambda  \left(2187 \lambda ^4+864 \lambda ^2+64\right)+18 \alpha ^5 \lambda ^2 \left(729 \lambda ^4-288 \lambda ^2-128\right)-3 \alpha ^4 \lambda ^3 \left(729 \lambda ^4-3456 \lambda ^2-640\right)
     \\+\alpha ^3 \left(480 \lambda ^4-4536 \lambda ^6\right)+72 \alpha ^2 \lambda ^5 \left(9 \lambda ^2-14\right)+384 \alpha  \lambda ^6-48 \lambda ^7,
 \end{multline}
 
 \begin{equation}
     \Xi=-(\lambda -2 \alpha )^6 \left(\alpha ^2 \left(27 \lambda ^2+8\right)+4 \alpha  \lambda -4 \lambda ^2\right)^3 \left(32 \alpha ^3 \lambda -\alpha ^2 \left(161 \lambda ^2+24\right)+4 \alpha  \lambda  \left(8 \lambda ^2-3\right)+12 \lambda ^2\right).
 \end{equation}
  \twocolumngrid
 \par 
 For the B critical point we have analyzed the values of the corresponding eigenvalues and the associated dynamical behavior. In Fig.~\ref{fig:punctbeiv} we have considered an interpolation method based on a neural network approach, determining a possible region where the $B$ cosmological solution has a stable dynamical behavior. Note that the corresponding dynamical behavior is influenced by the strength of the coupling with the cubic component (which encodes geometrical effects due to the third order contractions of the Riemann tensor), and the steepness of the potential energy.

\section{\label{sec:level4aa} Beyond the exponential potential}
 
\par 
In this section we shall investigate the structure and properties of the phase space in the case where the potential energy term is beyond the usual exponential case. To this regard, we shall consider that the potential is represented by an inverse hyperbolic sine, 

\begin{equation}
    V(\phi)=V_0 sinh^{-\xi}(\chi \phi),
\end{equation}
where $V_0, \xi, \chi$ are constant parameters. This potential has been considered in various cosmological models \cite{Roy:2017uvr, Bahamonde:2019gjk, Urena-Lopez:2000ewq, Sahni:1999gb}, representing a viable function. In order to study such a potential energy, we need to introduce another variable,
\begin{equation}
    \lambda=-\frac{1}{V(\phi)} \frac{dV(\phi)}{d \phi}.
\end{equation}
Then, the dynamical system is a four dimensional system, where we have to add the following differential equation, 
\begin{equation}
    \lambda'=-\sqrt{3} \lambda^2 x y (\Gamma-1),
\end{equation}
where $\Gamma$ is defined as follows:
\begin{equation}
    \Gamma=\frac{V(\phi) \frac{d^2 V(\phi)}{d \phi^2}}{(\frac{dV(\phi)}{d \phi})^2}.
\end{equation}
Taking into account the inverse hyperbolic sine decomposition, we have:
\begin{equation}
    \Gamma=1+\frac{1}{\xi}-\frac{\xi \chi^2}{\lambda^2},
\end{equation}
and the last dynamical equation reduces to:
\begin{equation}
    \lambda'=-x y \sqrt{3} (-\xi \chi^2+\frac{\lambda^2}{\xi}).
\end{equation}
In what follows we shall present the corresponding critical points and the specific fundamental properties obtained in the case where the potential is represented by a hyperbolic sine function. The first critical point found in our analysis is located at the following coordinates:
\begin{equation}
    C=\Big[x=0, y, z=\frac{1}{2}-\frac{y^2}{2}, \lambda=\frac{2 \alpha (y^2-1)}{y^2}
    \Big].
\end{equation}
We can note that for this solution the field is at rest, without any kinetic energy, while the  $y$ variable associated to the potential energy represents an independent parameter. The specific value of the last two variables $(z, \lambda)$, which are associated to the non--minimal coupling function and the steepness of the potential, respectively, are influenced by the potential component. From a physical point of view this cosmological solution describes a de--Sitter epoch $(\Omega_m=0, w_{\bf{eff}}=-1)$. For this solution, the general form of the eigenvalues is too complex to be written here. However, if we set $y=1$, we obtain a simpler form of the corresponding eigenvalues:
\onecolumngrid
\begin{equation}
    \Big[0,3 \left(w_m+1\right),\frac{1}{2} \left(-\sqrt{12 \xi  \chi ^2+9}-3\right),\frac{1}{2} \left(\sqrt{12 \xi  \chi ^2+9}-3\right)  \Big],
\end{equation}
\twocolumngrid
describing a non hyperbolic solution which is saddle. In this case the dynamical aspects are influenced by the values of the $\xi$ and $\chi$ coefficients. 
\par 
The second cosmological solution is represented by the following critical point, 
\begin{equation}
    D^{\pm}=\Big[x=\pm \sqrt{1+w_m}, y=\frac{\sqrt{3} \sqrt{w_m+1}}{\xi  \chi }, z=0, \lambda=\xi  \chi \Big]. 
\end{equation}
This solution corresponds to a matter dominated era $(\Omega_m=1-\frac{3 \left(w_m+1\right)}{\xi ^2 \chi ^2 \sqrt{-w_m}}, w_{\bf{eff}}=w_m)$, having the following eigenvalues (for the $D^{+}$ solution):
\onecolumngrid
\begin{multline}
    \Bigg[-\frac{6 \left(w_m+1\right)}{\xi },\frac{3 \left(w_m+1\right) (\alpha -2 \xi  \chi )}{\xi  \chi },
    \\ \frac{3}{4} \left(\pm \frac{\sqrt{\xi ^6 \chi ^4 w_m^2 \left(w_m+1\right) \left(17 \xi ^2 \chi ^2 w_m^2+14 \xi ^2 \chi ^2 w_m+48 \left(-w_m\right){}^{5/2}+48 \sqrt{-w_m}+96 \sqrt{-w_m} w_m+\xi ^2 \chi ^2\right)}}{\xi ^4 \chi ^3 w_m \sqrt{w_m+1}}+w_m-1\right)  \Bigg].
\end{multline}
\twocolumngrid
Hence, this physical point is similar to the $B$ solution discussed earlier in the case where the potential is represented by an exponential function. From a dynamical perspective if we consider that the dark matter pressure is negative but very close to zero and $\xi < 0$, then the first eigenvalue becomes positive, implying that this solution cannot be stable. Hence, it is either saddle or unstable, depending of the values of the specific coefficients $\{\alpha, \xi, \chi\}$. By imposing that the second eigenvalue is negative and $w_m=-0.0001$, we have obtained some constraints to the specific coefficients, 
\onecolumngrid
\begin{equation}
   \xi <0\land \left(\chi <0\lor \alpha >0\lor \chi >\frac{0.5 \alpha }{\xi }\right)\land \left(\chi <\frac{0.5 \alpha }{\xi }\lor \alpha \leq 0\lor \chi >0\right),
\end{equation}
\twocolumngrid
where the matter dominated era appears as a saddle cosmological solution in the phase space structure. As can be noted, the structure of the phase space in the case where the potential energy term is represented by an inverse hyperbolic sine is similar to the exponential potential case. 

\section{\label{sec:level5}Summary and Conclusions}

In the present manuscript we have proposed a novel cosmological model, by adding to the Einstein--Hilbert Lagrangian a tachyonic field non--minimally coupled with a topological invariant constructed with specific third order contractions of the Riemann tensor. In this scenario the dark energy component is represented by the tachyonic field which depends on the cosmic time. After proposing the action corresponding to this cosmological system, we have obtained the basic equations which describe the evolution of such a theoretical model, obtained by applying the variational principle. To this regard, the Klein--Gordon equation is obtained by varying the action with respect to the tachyonic field, while the modified Friedmann relations are deduced by varying the inverse metric. Since the present model does not take into account an interaction between the tachyonic field and the matter component, the standard continuity equation is also satisfied.
\par 
The physical features of the current cosmological system are investigated by adopting the dynamical system analysis, an important tool used in the study of various modified gravity theories. In the present paper we have analyzed the phase space structure and properties in the case of an exponential coupling function. Furthermore, for the potential energy we have considered two specific cases, the exponential potential and a distinct potential, the hyperbolic potential. In the case of an exponential potential the structure of the phase space has three dimensions, having two types of cosmological solutions. The first type is represented by the de--Sitter epoch, a cosmological solution where the tachyonic field acts as a cosmological constant, with a constant equation of state. At this critical point the value of the coupling coefficient which encodes specific interactions with the topological cubic invariant is affecting the dynamical consequences. 
\par 
To this regard, we have obtained specific constraints for the coupling coefficients where the attractor  behavior is attained in the distant future. The second type of cosmological solutions is represented by the matter dominated era, an epoch where the total (effective) equation of state of the cosmological system corresponds to the matter component. From a physical point of view this solution is viable only if we take into account that the dark matter fluid has a negative equation of state, slightly close to zero. This implies that the dark matter component corresponds to an exotic fluid with a negative pressure. From a theoretical perspective such a solution has been also found in different tachyonic dark energy models \cite{Bahamonde:2019gjk}. In the case of the second solution the dynamical features have been investigated, revealing some values of the coupling coefficients where the matter epoch corresponds to a saddle behavior, compatible with the recent evolution at the large scale structure.
Lastly, we have also considered that the potential energy term is represented by an inverse hyperbolic sine function, discussing the phase space structure and the corresponding dynamical effects. Finally, due to the presented arguments we can note that the present cosmological setup represents a viable alternative theory which can explain the evolution of the Universe -- the matter dominated epoch and the dark energy phenomenon, constituting a feasible theoretical framework, at least at the level of background dynamics.

\begin{acknowledgments}
This work was supported by a grant of the Romanian Ministry of Research, Innovation and Digitalization, CNCS - UEFISCDI, project number PN-III-P4-ID-PCE-2020-1142, within PNCDI III. For this work we have considered various computations in \textit{Wolfram Mathematica} \cite{Mathematica} and \textit{xAct} \cite{xact}.
\end{acknowledgments}

\bibliography{apssamp}

\end{document}